# Analysis of the planetary boundary layer with a database of large-eddy simulation experiments

**I. Esau[a,b]**

[a] G.C. Rieber Climate Institute at the Nansen Environmental and Remote Sensing Center, Thormohlensgt. 47, 5006, Bergen, Norway

[b] Bjerknes Center for Climate Research, Bergen, Norway

Corresponding author: Tel. +47 55205876, E-mail address: igore@nersc.no (Igor Esau)

## Abstract

Observational studies of a planetary boundary layer (PBL) are difficult. Ground-born measurements usually characterize only a small portion of the PBL immediately above the surface. Air-born measurements cannot be obtained close to the surface and therefore cannot capture any significant portion of the PBL interior. Moreover, observations are limited in choice of instrumentation, time, duration, location of measurements and occasional weather conditions. Although turbulence-resolving simulations with a large-eddy simulation (LES) code do not supplant observational studies, they provide valuable complementary information on different aspect of the PBL dynamics, which otherwise difficult to acquire. These circumstances motivated development of a medium-resolution database (DATABASE64) of turbulence-resolving simulations, which is available on ftp://ftp.nersc.no/igor/. DATABASE64 covers a range of physical parameters typical for the barotropic SBL over a homogeneous rough surface. LES runs in DATABASE64 simulate 16 hours' evolution of the PBL turbulence. They are utilized to study both transition and equilibrium SBL cases as well as to calibrate turbulence parameterizations of meteorological models. The data can be also used to falsify theoretical constructions with regards to the PBL.

## 1  Introduction

The turbulent planetary boundary layer (PBL) is often stably stratified, e.g. at high latitudes and in night time at all latitudes. A stably stratified PBL (SBL) is characterized by a lower surface temperature relative to temperature of air above it. The principal SBL feature is the downward, toward the surface, direction of the mean vertical sensible heat flux. There are at least three physical mechanisms leading to the SBL development, namely, the warm air advection, the radiative heat loss due to larger thermal emissivity of the surface, and the elevated latent heat release in clouds. Those mechanisms often interplay to create quite complicated structures of the SBL. The exact re-stratification mechanisms are however out of focus in this work.

We consider the PBL re-stratification as an external factor. Hence we focus on the SBL structure consistent with given external parameters within a typically observed range. Numerical large-eddy simulation (LES) technique is a method to generate the SBL structure consistent with specified and controlled external parameters. The external parameters are: the time-independent downward temperature flux, $F_\theta = \overline{w'\theta'} = H_s / \rho c_p$ at the surface ($z = 0$), and the initially imposed temperature stratification, $N = \left(g \theta_0^{-1} \partial \theta / \partial z\right)^{1/2}$. The former, $F_\theta < 0$ in SBL, is a prescribed constant, which is more typical for the surface with large heat capacity, e.g. water surfaces. The later, $N$ or the Brunt-Vaisala frequency, is modified within the SBL by the turbulent mixing. There is an interest to follow its evolution. Here, $\rho$ is the air density, $c_p$ is the air specific heat at the constant pressure, $H_s$ is the sensible heat flux, $z$ is the height above the impenetrable surface, $g$ is the gravity acceleration, $\theta_0$ is the reference temperature characterizing the air thermal expansion, $\theta$ is the potential temperature.

Preparing the SBL simulations, it is perhaps the most difficult to compose a range of its characteristic external parameters. The SBL is scarcely known. Esau and Sorokina (2009) composed the SBL climatology for the Arctic on the basis of literature review and data processing. This climatology is used to select relevant LES runs for the database. Additional runs were motivated by the needs to improve SBL parameterisations (Mahrt, 1998; Tjernstroem et al., 2005; Cuxart et al., 2006; Beare et al., 2006; Holtslag et al., 2007; Steeneveld et al., 2008; Byrkjedal et al., 2008). Mauritzen et al. (2007) have already used this database for such a crucial improvement.

To understand the added value of the LES database, one should consider the best traditional data obtained during short field campaigns. The SBL observations are difficult. On the one side, ground-born measurements usually characterize only a small portion of the SBL immediately above the surface and therefore strongly affected by local surface features. The typical height of a meteorological mast is about 10 m to 30 m, whereas even the shallowest SBL are of 50 m to 100 m deep. Moreover, the SBL is often capped with temperature inversion. The inversion in potential temperature is observed in more than 95% of the cases in Northern high latitudes (Serreze et al., 1992; Liu et al., 2006). On the other side, the air-born measurements usually cannot be done sufficiently close to the surface to capture any significant portion of the SBL. Flights below 50 m level are problematic, although development of unmanned aircrafts can facilitate the low-level data sampling (Reuder et al., 2008). Recent advances in remote sensing techniques (Bösenberg and Linné, 2002; Emeis et al., 2004; Cooper et al., 2006; Frehlich et al., 2008), dropsonde and tethered balloon techniques can provide high quality data (Holden et al., 2000) but require considerable improvement of resolution, especially near the surface. In these circumstances, LES can provide a valuable supplement data for complementary and independent studies of the SBL structure and dynamics (Stevens and Lenshow, 2001).

In some ways, LES data are indispensible as they can provide information, which is difficult if possible to acquire by other approaches. In particular, LES is useful due to easy accessibility, controllability, accuracy and repeatability of its data. The LES can be rerun at any request to check for impact of different physical and numerical factors as well as to obtain any kind of dynamical quantity of interest. The major LES drawbacks are related to unavoidable idealizations in model experiments. Therefore, it is important not to overstate the limits of the LES data validity. As it is recognized now (Mason, 1994; Muschinski, 1995; Esau, 2004), the LES data characterise dynamics of a LES-fluid flow, which in many aspects resembles the atmospheric flow, but may differ from it in aspects ultimately sensitive to the properties of the smallest and the largest resolved-scale motions in the model. Hence, the LES data must be always falsified against observations where the observations are available.

This paper describes a database (DATABASE64) of turbulence-resolving simulations. DATABASE64 covers the physical parameters characterizing the barotropic SBL over a homogeneous aerodynamically rough surface. Each LES run in DATABASE64 simulates 16 hours' evolution of the SBL. The LES produces three dimensional fields of fluctuations of the meteorological variables, namely, the three components of velocity and the potential temperature. These fluctuations are further processed to obtain turbulence statistics as well as horizontally and time averaged profiles of the required quantities. DATABASE64 can be used for both transition and equilibrium SBL studies as well as for calibration of the turbulence parameterizations. DATABASE64 is available on ftp://ftp.nersc.no/igor/ and it has been already used in several research studies. Esau and Zilitinkevich (2006) utilized DATABASE64 to characterize the SBL parameter space in order to develop analytical resistance laws. Zilitinkevich et al. (2007; 2008) used DATABASE64 to support a total turbulent energy theory. Mauritzen et al. (2007) and Canuto et al. (2008) utilized DATABASE64 to calibrate new parameterizations. Several other research groups have also expressed their interest in DATABASE64 (e.g. Frehlich et al., 2008; Basu et al., 2008; Jeričević et al., 2009; Perez et al., 2009; Grisogono and Rajak, 2009).

This study aims to demonstrate several new ways to work with DATABASE64. The examples make contribution to selected problems of the boundary layer theory, parameterisations and understanding. The paper is organized in the following way. The next Section describes the numerics of the large-eddy simulation code LESNIC (Esau, 2004). The attention is paid to treatment of fluid static stability. This Section is given to facilitate the understanding of the LES advantages and limitations. The reader may omit it if those details are of no interest for him. Section 3 describes the numerical experiments and results of sensitivity tests. Section 4 gives several examples of DATABASE64 applications. Section 5 summarizes the presentation.

## 2   The large-eddy simulation code LESNIC

The present database has been obtained with the large-eddy simulation code LESNIC developed by the author at Uppsala University and Nansen Environmental and Remote Sensing Center (Esau, 2004). The code numerically solves Navier-Stokes equations of motions for incompressible Boussinesq fluid and the transport equations for the potential temperature and passive scalars. The equations for passive scalars are identical to that for the potential temperature and will not be described here.

Let us define the Cartesian coordinate system $(x, y, z)$ with the axis directed to East, North and Zenith correspondingly. The components of velocity in this coordinate system are $u_i = (u, v, w)$. The potential temperature is defined as $\theta = T(p_0/p)^{0.287}$ where $T$ is the absolute temperature and $p$ is the pressure with $p(z=0) = p_0$. The layer of neutral static stability would have $\partial \theta / \partial z = 0$. Equations of motions, continuity and the scalar transport are

$$\partial u_i / \partial t = -\partial / \partial x_j (u_i u_j + \tau_{ij} + p\delta_{ij}) + F_u, \tag{1}$$

$$\partial u_i / \partial x_i = 0, \tag{2}$$

$$\partial \theta / \partial t = -\partial / \partial x_j (\theta u_j + \tau_{\theta j}) + F_b. \tag{3}$$

Here we use the Einstein rule of summation and the Kroneker delta, which is $\delta_{ij} = 0$ if $i \neq j$ and $\delta_{ij} = 1$ otherwise. In Eq. (1)-(3), the Boussinesq approximation has been applied (Zeytounian, 2003). The pressure $p$ must be understood as the deviation from the hydrostatic pressure. The continuity equation reduces to the non-divergence equation by this approximation.

The external forces specified in this set of simulations are the three dimensional Coriolis and reduced buoyancy forces. The forces are $F_u = F_{coriolis} + F_{buoyancy}$ where $F_{coriolis} = -2(\vec{\Omega} \times \vec{u})$ and $F_{buoyancy} = -g\theta_0^{-1}\theta\delta_{i3}$. We take and $F_\theta = 0$ in DATABASE64.

The terms, $\tau_{ij}$ and $\tau_{\theta j}$, are of special interest. They are responsible for the energy dissipation and the temperature fluctuations' diffusion in the LES code. Therefore, modeling of $\tau_{ij}$ and $\tau_{\theta j}$ is critical for the LES performance and accuracy. In LESNIC, $\tau_{ij}$ and $\tau_{\theta j}$ are not dissipative terms as it is adopted in turbulence parameterizations in the large-scale meteorological models. The terms $\tau_{ij}$ and $\tau_{\theta j}$ have the same mathematical properties as the resolved transport terms, $u_i u_j$ and $\theta u_j$, but describe the spectral energy (temperature fluctuation) transport across the smallest resolved scale (the mesh scale) in the code. They are traceless tensors with 6 independent components for $\tau_{ij}$ and 3 components for $\tau_{\theta j}$.

It should to be clearly understood that $\tau_{ij}$ and $\tau_{\theta j}$ cannot be constructed by analogy with the molecular dissipation/diffusion as it is done in the meteorological parameterizations. To construct $\tau_{ij}$ and $\tau_{\theta j}$, LESNIC utilizes an analytical solution of a simplified variational optimization problem for the spectral energy transport in the inertial interval of scales. The detailed derivation, tests and proper references can be found in Esau (2004). The essence of this variational problem (Pope, 2004) is to find such values of the term $\tau_{ij}$ that balance as accurately as possible the amount of energy cascading through the mesh scale, $\Delta$, with the amount of energy cascading through some larger resolved scale, $\Delta^L > \Delta$. The latter value can be explicitly computed, as the term $L_{ij}^L$ (see Eq. (6) below), up to the accuracy of the numerical scheme. In a nearly laminar low Reynolds number (Re) flow, where the direct interaction hypothesis[1] is justified, invoking of $L_{ij}^L$ into the formulation for $\tau_{ij}$ provides almost exact spectral turbulence closure. Unfortunately, in the high Re flow such as the atmospheric SBL, a large fraction of energy is cascading indirectly through interactions between motions with significantly different scales. In this case, $L_{ij}^L$ is not sufficient to describe the magnitude of the spectral energy transport. $L_{ij}^L$ saturates at about 50% of the total turbulent stress magnitude on a fine resolution mesh (Sullivan et al., 2003). More sophisticated constructions for $\tau_{ij}$ are required (Vreman et al., 1997).

Depending on the choice of optimization methods and amount of information extracted from the resolved flow fluctuations[2] different turbulence models were proposed. LESNIC employs relatively unsophisticated and therefore computationally inexpensive model. It is a reduced version of a dynamic-mixed model that reads

$$\tau_{ij} = L_{ij}^L - 2l_s^2 |S_{ij}| S_{ij}, \tag{4}$$

$$l_s^2 = \frac{1}{2} \frac{(L_{ij}^L - H_{ij}^L) \bullet M_{ij}^L}{M_{ij}^L \bullet M_{ij}^L}, \tag{5}$$

$$L_{ij}^L = (u_i u_j)^L - (u_i)^L (u_j)^L, \tag{6}$$

$$H_{ij}^L = ((u_i)^L (u_j)^L)^L - ((u_i)^L)^L ((u_j)^L)^L - [((u_i)')^L (u_j)')^L - ((u_i)'(u_j)')^L], \tag{7}$$

$$M_{ij}^L = (|S_{ij}| S_{ij})^L - \alpha |(S_{ij})^L|(S_{ij})^L, \tag{8}$$

$$S_{ij} = \frac{1}{2}(\partial u_i / \partial x_j + \partial u_j / \partial x_i). \tag{9}$$

---

[1] The most energy exchange is due to the interaction between eddies with the closest wave lengths.

[2] Following Germano (1986), Carati et al. (2001) showed that the exact expression of $\tau_{ij}$ through the resolved-scale fluctuations of the velocity could be obtained with the infinite Taylor series expansion. The more terms of the expansion are retained the more accurate and complete $\tau_{ij}$ is obtained. The computational cost of this approach increases rapidly, rendering it of being mostly of the academic interest.

Here $A_i \bullet A_j$ is the scalar product, $|A_i| = (A_i \bullet A_i)^{1/2}$, the superscripts $l$ and $L$ denote filtering with the mesh length scale and the twice mesh length scale filters. The filters' squared aspect ratio is $\alpha = 2.92$ for the Gaussian and the top-hat filters discretized with the 2$^{nd}$ order of accuracy central-difference schemes. The reader should observe that the formulation in Eq. (5) for the mixing length scale is given in quadratic form. It implies imagery values for the mixing length in certain flow conditions that physically means the turbulence energy backscatter from the small to large scales. The turbulence model for the temperature fluctuations is

$$\tau_{\theta j} = -\text{Pr}^{-1} l_s^2 |S_{ij}| (\partial \theta / \partial x_j), \text{ where} \tag{10}$$

$$\text{Pr} = \text{Pr}_0 + aRi, \quad \text{Pr}_0 = \lim_{Ri \to 0} \text{Pr} = 0.8, \quad a = 5, \text{ and} \tag{11}$$

$$Ri = g\theta_0^{-1} \partial \theta / \partial z \left(\partial |u_i| / \partial z\right)^{-2}. \tag{12}$$

The numerical discritization in the LESNIC code is the 2$^{nd}$ order fully conservative finite-difference skew-symmetric scheme on the uniform staggered C-type mesh with the explicit Runge-Kutta 4$^{th}$ order time scheme. The schemes and the relevant references are given in Esau (2004). The mesh scales are $\Delta_x, \Delta_y$ and $\Delta_z$ with the 1$^{st}$ computational level for $u, v, \theta$ is placed at $z_1 = \Delta_z / 2$ and for $w$ at $\Delta_z$. The lateral boundary conditions are periodic. It allows implementation of exact (up to the computer accuracy) and fast direct algorithms to solve the continuity Eq (2) with a pressure correction. But it also limits the size of resolved perturbations and spans of trajectories to ½ of the size of the computational domain and may cause an artificial energy accumulation at the longest resolved scales.

## 3  The experiment set up and quality assessment

DATABASE64 is a collection of LESNIC runs each computed independently from slightly perturbed laminar flow under a fixed set of external control parameters. The surface boundary conditions for temperature are prescribed in each run through a constant temperature flux

$$\tau_{\theta j}(t, z = 0) = const \tag{13}$$

The constant flux conditions have some drawbacks. Derbyshire (1999) and Basu et al. (2008) argued that the prescribed heat flux boundary conditions should be avoided in the SBL as there is a possibility to exceed the maximum physical heat flux consistent with the intensity of the flow turbulence. In this case, there is a possibility for runaway surface cooling in the LES.

Indeed, the simulations run into troubles under some parameter combinations crossing probably into unphysical areas of the parameter space (see examples in the next Section). Moreover, the bifurcations in the SBL turbulence dynamics cannot be simulated[3]. The surface boundary conditions for momentum are given through the log-layer formulation for the friction velocity

$$u_* = \kappa |u_i(z_1)| / \ln(z_1 / z_0), \tag{14}$$

$$\tau_{ij}(z = 0) = \delta_{i3} u_*^2 u_i(z_1) / |u_i(z_1)|. \tag{15}$$

The log-linear Monin-Obukhov stability functions are not used in LESNIC. Nevertheless, the consistent log-linear behaviour is recovered in the simulated fields.

The LES runs were integrated for 16 hours. The three dimensional data were sampled every 600 s and processed during the simulations. DATABASE64 contains the horizontally and time averaged data and turbulent statistics. The averaging was done over each half an hour of simulations that is over 3 subsequent samples. At the end of each run, the instant three dimensional fields of $u, v, w, \theta$ were stored in the database.

The turbulence statistics were obtained in the post-processing as follows. Consider for example the mean resolved vertical temperature flux profile $\overline{w'\theta'}$. Let $<\phi>_x = \frac{1}{n}\sum_{i=1}^{n}\phi(x_i, y, z, t)$ be an averaging operator in the direction $x$ where $n = L_x / \Delta_x$ is the total number of samples (grid nodes) along $x$. Then the flux is defined as

$$\overline{w'\theta'}(z, t') = << w(x, y, z, t) \cdot (\theta(x, y, z, t) - << \theta(x, y, z, t) >_x >_y) >_x >_y >_{t \in [t_i + \delta t]} \tag{16}$$

where $\delta t$ is the 3 subsequent instant data samples.

---

[3] Possibilities for sharp transitions in the intensity of the vertical turbulent mixing in the SBL were theoretically discovered by McNider et al. (1995). The transitions were linked to the bifurcations and hysteresis found in the SBL parameter space where two different states with low and high levels of the turbulent mixing are possible for the same range of control parameters. Derbyshire (1999) demonstrated the bifurcations in the SBL parameterizations. This kind of the behaviour has been supposedly reported by Lyons and Steedman (1981) for the Australian nocturnal SBL. It remains to be seen whether the LES also reproduce the SBL bifurcation dynamics.

The mesh size of 64 by 64 by 64 grid points is only marginally sufficient for the SBL studies (Esau and Zilitinkevich, 2006). The larger mesh size however would seriously increase the demand for the computational resources. For instance, the use of 128 by 128 by 128 mesh will increase the computer time per run by more than an order of magnitude. To assess the database quality several additional runs were conducted. The sensitivity of the LES runs to the mesh resolution has been shown in Esau and Zilitinkevich (2006). We will not repeat it here. More results on the LES sensitivity to the mesh resolution could be found in Beare and McVean (2001).

A lack of resolution severely damages the inertial interval (the large wave numbers) in the turbulence energy spectrum. The model partially absorbs the damage reducing the effective Smagorinsky constant $C_s = l_s / \Delta$ and hence the sub-grid scale flux $\tau_{ij}$. The averaged $C_s$ in the SBL core (from $z = h/3$ to $z = 2h/3$) drops from its nearly theoretical value (Leslie and Quarini, 1979) of $C_s = 0.2$ in the well resolved runs with 32 – 48 levels within the SBL to $C_s = 0.1$ for the runs with 12 – 24 levels within the SBL. The similar reduction of the grid-scale dissipation, and $C_s$, has been documented in the Horizontal Array Turbulence Study (HATS) atmospheric experiment (Kleissl et al., 2006). The dynamic $C_s$ reduction has limits of adjustment. It should not draw the average $C_s$ below ~0.1. Hence, more stable runs must be produced in smaller size domains. Here, the domain height varies from $L_z$ ~ 4000 m in the EBL to $L_z$ ~ 80 m in the most stable SBL. Without a careful intercomparison study it still remains unknown whether the domain size is adequate. The assessment was made only for one EBL run. Figure 1 shows that the turbulent statistics in DATABASE64 are not visibly sensitive to the horizontal dimensions of the domain. Here, the streamwise dimension $x$ varies from 4 km to 16 km or from about $2h$ to $8h$. The turbulence profile modifications due to the too restrictive size of the vertical domain dimension $z$ have been shown in Esau (2004) for the EBL. The main feature is the turbulence intensification in the boundary layer core and to some degree the intensification of the surface layer turbulence. The reason for the noted changes is a larger wind shear in the shallower boundary layer. Figure 2 shows the same effect for the nocturnal stable (NS) SBL runs. The restrictions on $L_z$ result in turbulence intensification in the SBL core, which lead to the stability reduction in further turbulence intensification. This effect is observed in the given example.

DATABASE64 is to characterize the equilibrium steady state SBL and to support the asymptotic calibration of parameterizations. Several extra-long runs have been conducted to assess the degree of the equilibrium and statistical steady state reached in the runs. Figure 3 shows this assessment. For three turbulence statistics considered, the equilibrium is reached by $3 tf$. The statistics fluctuates around the respective steady state values with the amplitudes of about $\pm 10\%$ but no regular trends can be observed up to $23 tf$ units. The reason for the fluctuations needs an additional investigation, which is out of scope here.

DATABASE64 has been calculated with a prescribed turbulent surface temperature flux $F_\theta$ in the boundary conditions. This approach has two advantages. First, it eliminates the need for the surface energy balance model, which could be a challenge for its own. Second, it provides the surface stability parameter as the external control parameter in the simulations. The approach has also important drawbacks. Derbyshire (1999) disclosed that the downward sensible heat flux cannot be arbitrary large and that for each given wind speed there is a certain finite turbulent flux to be supported by the flow. More careful study by McNider et al. (1995) and Derbyshire (1999) indicated that non-unique values of the flux is possible for the given wind speed in some observable range of values. Later, Basu et al. (2008) derived the criterion to analyse a posteriori consistency of the surface fluxes. The criterion has been applied to DATABASE64 to separate the physically consistent runs. Figure 4 shows an example of this work. Several runs, especially those with the strongest surface stability conditions, were recognized as inconsistent and deleted from DATABASE64. It noticeably improved the DATABASE64 quality through the elimination of the unexplainably scattered results of those inconsistent runs.

## 4  Study of the SBL utilizing DATABASE64

There are several ways to utilize DATABASE64. The analysis of the SBL structure can be done using data interpolation on the levels of a standard fraction of the SBL depth $z/h$. This analysis is exemplified in Section 4.1. The data can also be analysed in the parameter space (e.g. Esau and Grachev, 2007). This analysis is exemplified in Section 4.2. DATABASE64 can be also applied for non-stationary analysis of the SBL evolution in its transition to the equilibrium state as exemplified in Section 4.3. Finally, Section 4.4 presents slices of instant three-dimensional fluctuation fields for some runs and gives an example of using it to recover parameters, which are not included in the parameter list in DATABASE64.

## 4.1 Equilibrium vertical structure of the SBL

After initial spin up and following development, the simulated SBL approximately reaches the statistical equilibrium steady state by 13$^{th}$ through 16$^{th}$ hours of simulation. To obtain the vertical structure of the SBL, the quantity of interest is normalized and linearly interpolated on $z/h$-levels placed with the step $0.05h$. The interpolated values were additionally time averaged over the 3 hours' interval. Finally, the obtained profiles for each run were averaged for all runs in DATABASE64. At this stage, the standard deviation among the runs was computed. It corroborates that the mean profile is a good approximation of individual ones.

The equilibrium SBL depths, $h$, are given in Figure 5. The theory constructed with assistance from these data has been published in Zilitinkevich et al. (2007). The vertical momentum flux exhibits remarkable universality with relatively small scatter in Figure 6. The universal profile was found for the turbulent kinetic energy, $E = 1/2(\sigma_u^2 + \sigma_v^2 + \sigma_w^2)$ where $\sigma_{u_i}^2 = \overline{u_i' u_i'}$, and the variations of the vertical velocity component, $\sigma_w^2$, profiles in Figures 7 and 8. These results support the well established inner scaling (the scaling is based on the surface friction velocity, $u_* = |\overline{u_i' w'}|^{1/2}, i = 1,2$).

The thermal (density) SBL structure is more complicated. There are two sources of stability in DATABASE64, namely, the applied constant surface temperature flux, $F_\theta$, at $z = 0$, and the initially prescribed temperature stratification, $N(t = 0)$, of the flow. Each source creates own universal asymptote in the nocturnal (NS) SBL and the conventionally (CN) SBL (Zilitinkevich and Esau, 2003) correspondingly. Figure 9 shows that the NS SBL universality is well established, whereas the CN SBL universality, and therefore the universality in the long-lived (LS) SBL, is much less so. The cause of scatter in this case is development of the capping inversion. This development has been shown by Nieuwstadt (1984) in his analytical SBL model. The stratification at the SBL top gradually increases with time making the top-down, $F_{\theta\downarrow}$, component of the temperature flux comparable to the traditional bottom up, $F_{\theta\uparrow}$, component. On average, although the vertical velocity fluctuations are small at the SBL top, $N$ can be very large[4]. It results in large values of the turbulent potential energy, $E_p = 2\overline{\theta'\theta'}(g\theta_0^{-1}N^{-1})^2$ (Zilitinkevich et al., 2008), as seen in Figure 10. In some runs it may temporally lead to the height-independent total temperature flux $F_\theta = F_{\theta\downarrow} + F_{\theta\uparrow}$. This may partially explain failure of the meso-scale meteorological models in nowcast and prediction of the optical turbulence (Lascaux et al., 2009), which is critically depend on $F_{\theta\downarrow}$ unaccounted in the existing parameterizations.

---

[4] Some studies, e.g. Balsley et al. (2003), reported the observed temperature gradient up to 20 K m$^{-1}$, which corresponds to $N \sim 0.8$ s$^{-1}$ as compared to the typical $N \sim 0.01$ s$^{-1}$.

## 4.2 Equilibrium parametric structure of the SBL

The parametric dependences define relationships between resolved and unresolved variables in models. Therefore they are import for development of parameterizations. The initial and boundary conditions prescribe the values of the external parameters to control each LES run. With variety of the LES runs, DATABASE64 provides a range of parameter values. The inter-run analysis could help extracting the parametric dependences valid within this range. As in Section 5.1, we use the equilibrium quasi-steady state interpolated data. One example is seeking for the Monin-Obukhov stability functions for the non-dimensional velocity, $\Phi_M$, and temperature, $\Phi_H$, gradients. These functions play the fundamental role in the parameterizations (e.g. Zilitinkevich and Esau, 2007; Basu et al., 2008). A generalized theory of these functions is given in Zilitinkevich et al. (2009). The functions read

$$\Phi_M = \kappa z \tau^{-1/2} \partial U / \partial z,  \qquad (17)$$

$$\Phi_H = \kappa_T z \tau^{1/2} F_\theta^{-1} \partial \theta / \partial z,  \qquad (18)$$

where $\kappa = \kappa_T = 0.43$ is the von Karman constant defined in post-processing through the best fit of the $\Phi_M(z)$ from the truly neutral boundary layer case in DATABASE64 to the theoretical profile $\Phi_M = 1$. The inter-run averaged $\Phi_M$ and $\Phi_H$ are shown in Figures 11 and 12 as functions of the parameter $\varsigma = z / L_*$. The parameter $\varsigma$ can be obtained iteratively or algebraically in the meteorological models in the case the Monin-Obukhov functions are known. Therefore, the utility of the LES is to provide those functions for the parameterization schemes. Here, the height scale $L_*$ is defined following (Zilitinkevich and Esau, 2005; 2007) as

$$1/L_* = \left( L^{-2} + \left( L_N / C_N \right)^{-2} + \left( L_f / C_f \right)^{-2} \right)^{1/2},  \qquad (19)$$

$$L_N = \tau^{1/2} / N \text{ and } L_N = \tau^{1/2} / |f|, \tag{20}$$

where $L = -u_*^3 \left(g\theta_0^{-1} F_\theta\right)^{-1}$ is the Monin-Obukhov length scale and empirical constants are $C_N = 0.16 \pm 0.1$ and $C_f = 1 \pm 0.1$.

It is worth to emphasize that each run in DATABASE64 spans only some sub-range of $\varsigma$. Thus, in order to obtain the universal stability functions over 4 decades of $\varsigma$ magnitude, the inter-run analysis must be applied. Several approaches are possible. Zilitinkevich and Esau (2007) applied the analysis in the following way. $\Phi_M(z,n)$, $\Phi_H(z,n)$, and $\varsigma(z,n)$ were computed at each level $z$ (different for each run) in each run $n$. The one hour averaging was applied for the last, 16$^{th}$ hour to the data. The obtained range of $\varsigma$ was split on $M$ bins, $m = 1...M$, and the subsequent inter-run averaging was applied to within each bin $\varsigma \in (\varsigma_m, \varsigma_{m+1})$. The theoretically expected dependence was then fitted to the obtained averaged values. This approach is applied in the other author's works as well. Another approach, as in the present study, first produces interpolation on standard levels $z/h$ for which $\varsigma(z/h,n)$ are computed for each run $n$. Then the inter-run averaging was applied. The obtained results were split on $M$ bins. The results are only marginally different from the previous approach.

### 4.3 Evolution of the SBL

The raw LES output allows tracking the evolution with 10 min or even better time resolution. DATABASE64 retains the time resolution of 30 min to improve the data quality. Thus, it resolves processes on the hour or longer time scales. There are two principal processes with such long time scales in the SBL, namely, the initial model spin-up and the inertial oscillation (e.g. Buajitti and Blackadar, 1957).

The initial model spin-up is unavoidable in simulations starting from arbitrary (e.g. laminar) initial conditions. Its nature is primary numerical. Hence its analysis requires a high degree of the simulation technique's understanding. We will not stop here on the spin-up analysis in details. The spin-up analysis could be utilized to understand perturbation growth and instabilities in the LES. The spin-up lasts about 1.5 *tf* or 2 to 3 hours. It is significantly shorter in the shallow SBL. During the spin-up, the initial random perturbations lost their energy in stable directions of the numerical problem eigenspace. Simultaneously they gain energy in the unstable directions. Thus the principal process is the structural modification of the flow from being perturbed randomly to being perturbed by the turbulence. In LESNIC, this process leads to initial decay of the perturbation energy following by exponential growth of the energy of the consistent turbulence perturbations. The energy of those earlier perturbations overshoots the physically consistent level as the energy cascade to the dissipation short scales has not been developed yet.

The spin-up process is illustrated here using the SBL depth, $h$, development. Figure 13 shows the normalized quantity $h(t)/h_{theory}$ where the theoretical values are given by Zilitinkevich et al. (2007) as

$$h_{theory} = \left( \frac{f^2}{(C_R u_*)^2} + \frac{N|f|}{(C_{CN} u_*)^2} + \frac{|f \beta g F_\theta|}{(C_{NS} u_*^2)^2} \right)^{-1/2}, \qquad (21)$$

where $C_R = 0.55 \pm 0.05$, $C_{CN} = 1.36 \pm 0.25$, $C_{NS} = 0.51 \pm 0.06$ are empirical constants. DATABASE64 gives $h$ defined from the turbulent momentum flux profile as the height where $\tau(z) < 0.05 \tau_0 = 0.05 u_*^2$. This definition is preferable as it distinct the SBL on the basis of the turbulent momentum flux. Regular meteorological data, e.g. radiosoundings, do not allow calculation following this definition. Therefore, DATABASE64 provides also $h$ defined through the bulk Richardson number as in Troen and Mahrt (1986). The limit $\lim_{t \to \infty} h(t)/h_{theory} \to 1$ should not surprise as the constants in $h_{theory}$ where calibrated with the same data. The spin-up process is mostly finished by 1.5 $tf$. Moreover, its amplitude is more limited in the shallow SBL.

Contrary to the spin-up process, the inertial oscillation (IO) nature is physical but it could be caused by unphysical initial conditions and the peculiarity of the spin-up process as well. The IO develops when the flow velocity is in the inertial imbalance with the acting forces in the rotated frame of reference. Figure 14 shows the wind hodographs at the SBL top in DATABASE64. The IO should not appear in the idealized simulations of continuously growing SBL. The IO emerges in DATABASE64 due to the initial overshooting in the spin-up process. As the turbulence decay at the upper levels after the spin-up, there is no dissipation without turbulence in the LESNIC code to suppress the IO.

### 4.4 Instant three-dimensional fields of fluctuations

Many research directions, where DATABASE64 potentially can be of utility, have not been foreseen. In order to make DATABASE64 more useful and flexible, instant three-dimensional fields of velocity component and temperature fluctuations were stored at the end of $16^{th}$ hour of simulations. For some selected runs much more raw information have been stored like 30 to 60 successive samples of the three-dimensional fields with time resolution of 1 to 10 min. These additions are not included in DATABASE64 but can be requested from the author.

Figure 15 presents the three-dimensional fields of anomalies, $u'(x,y,z) = u(x,y,z) - \langle\langle u(x,y,z)\rangle_x\rangle_y$, of the potential temperature and the horizontal component of velocity at $z = 0.1h$ for the most stable SBL where $h$ is the smallest. Those fields are not only interesting as such but they allow computing quantities that are not in DATABASE64. The lack of time averaging can be, to some degree, compensated by space averaging. Figure 16 illustrates this kind of analysis using the cross-wind spectra of the horizontal velocity at different heights within the SBL. The obtained spectra are further processed to extract the length scale of the spectral maximums at each level. This length scales are associated with spanwise size of coherent structures in the SBL (Jimenez, 1998). Figure 17 shows the vertical profile of this length scale averaged in the streamwise direction and its corresponding standard deviations computed from slices along $x$ direction. The analysis strongly suggests that the SBL possesses the longitudinal roll-like structures confined in the turbulent layer below $h$ but above the surface layer of $0.2h$.

### 5 Summary

The paper presents the study of the SBL with the LES database DATABASE64. The LES runs were conducted with the code LESNIC on the mesh with the resolution of 64 by 64 by 64 grid nodes in each direction. Each LES run was computed over $16^{th}$ model hours. The paper exemplifies several ways to utilize DATABASE64, namely, the use of the equilibrium steady state SBL characteristics, the DATABASE64 parameter space, the SBL transient evolution and the advanced research utilizing the raw instant data fields.

The quality of the DATABASE64 is assessed against fine resolution additional runs as well as theoretical considerations. In general, the quality is reasonable. The vertical resolution of 30 to 50 levels per the SBL is sufficient for the most applications since more than 90% of the energy of fluctuations is resolved (Meyers and Baelmans, 2004). It is shown that the essentially better quality could be obtained only with at least twice as fine grid that requires an order of magnitude more computer time (and the similar increase of the memory space) per each run.

## Acknowledgements

The work was supported over several year by the Norwegian Research Council projects PAACSIZ 178908/S30, POCAHONTAS 178345/S30, and NORCLIM 178245/S30, as well as by ongoing projects EU FP7 MEGAPOLI "Megacities: Emissions, urban, regional and Global Atmospheric POLlution and climate effects, and Integrated tools for assessment and mitigation" and FP7-IDEAS "Atmospheric planetary boundary layers: physics, modelling and role in Earth system", the Norwegian Research Council PBL-feedback 191516/V30.

**Supplementary material**

After quality control, about 60 runs compose DATABASE64 at present. The runs cover a large variety of the external control parameters and represent situations from the truly neutral Ekman boundary layer (EBL) to very stable SBL at different latitudes, with the geostrophic wind of different speed, ambient atmospheric stratifications, and surface roughnesses.

With this presentation, readers could find the MATLAB (-MAT) file with DATABASE64. The MATLAB data base of the three-dimensional instant snapshots at $16^{th}$ hour of simulation (about 300 Mb) can be provided by request through the ftp service. The complete data set, initial and boundary conditions and additional experiments to assess the data quality are stored in the Nansen Environmental and Remote Sensing Center, Bergen, Norway. DATABASE64 is continuously complemented with new LES runs and some experiments are rerun at better quality. The data base is gradually evolving to the finer resolution and better quality product. Therefore, it is recommended to check with the author for the latest version of the product.

The examples of MATLAB scripts to open and work with DATABASE64 are also supplemented with this paper. The supplements are listed in Table 1. Table 2 describe the variables' convention in DATABASE64.

Table 1. Supplemented files.

| File name | Type | Comments |
| --- | --- | --- |
| DATABASE64_16hr_final.bin | MATLAB – MAT file | The data base DATABASE64 |
| DATABASE64_16hr_final.m | MATLAB script | The control script to process the raw LES data into data base |
| getdata64s.m | MATLAB function | The function called from the control script to fill in the DATABASE64 fields |
| sbl_vertical_structure.m | MATLAB script | The example of the equilibrium steady state vertical profiles calculated from the SBL with inter-run averaging and interpolation on standard levels (Section 5.1) |
| sbl_vertical_structure _mean_profiles.m | MATLAB script | The example of the equilibrium steady state parametric dependences calculated from the SBL with inter-run averaging and interpolation on standard levels (Section 5.2) |
| sbl_vertical_structure _mean_profiles_theta.m | MATLAB script | |
| sbl_evolution_depth.m | MATLAB script | The example of the time evolution calculated from the SBL with inter-run averaging (Section 5.3) |
| sbl_3D_analysis.m | MATLAB script | The example of the three-dimensional field analysis calculated from one SBL run (Section 5.4) |

Table 2. Variables in the structure fields (e.g. *db.class*) in DATABASE64.

| Field name | Array Size | Description |
|---|---|---|
| class | NA | A priori experiment definition: tnt – truly neutral boundary layer ($N = 0$, $F_\theta = 0$); cnt – conventionally neutral boundary layer ($N > 0$, $F_\theta = 0$); tst – nocturnal SBL ($N = 0$, $F_\theta < 0$); cst – long-lived SBL ($N > 0$, $F_\theta < 0$). |
| comment | NA | Commentary |
| fcomp | NA | Computer type: b – IEEE floating point with big-endian byte ordering; l – IEEE floating point with little-endian byte ordering |
| fname | NA | Folder name where the raw data are placed |
| ndim | 7 | The run dimensions 1 through 7: $Nx, Ny, Nz$ - the number of grid points in $(x, y, z)$ directions; $Nt$ - the number of the instant samples (profiles) in the raw data; $Nv$ - the number of the pre-calculated turbulence statistics in the raw data; $Ns$ - the number of the pre-calculated surface parameters the raw data; $Ndt$ - the number of samples used for time averaging; |
| Lat | 1 | Latitude (degrees North) |
| Omega | 1 | The Earth's angular velocity (7.29 $10^{-5}$ s$^{-1}$) |
| f | 1 | The Coriolis parameter (s$^{-1}$) |
| z0 | 1 | The surface roughness length scale (m) |
| Ug | 1 | The module of the geostrophic wind force (m s$^{-1}$) |
| Ug_dir | 1 | The geostrophic wind force direction left of the latitude |
| Ro | 1 | The Rossby number |
| kappa | 1 | A priori (prescribed) von Karman constant |
| z | $Nz = 64$ | The height (m) above the impenetrable surface |
| dz | 1 | The distance between the adjacent vertical levels |
| caseTime | $NT = Nt / Ndt = 32$ | Time (s) of sampled and averaged data |
| us | $NT$ | The surface friction velocity (m s$^{-1}$), $u_*$ |
| us_dir | $NT$ | The direction (degrees) of the surface stress vector |
| ts | $NT$ | The surface temperature flux scale (K), $\theta_*$ |
| wt0 | $NT$ | The surface temperature flux (K m s$^{-1}$), $F_\theta$ |
| Ls | $NT$ | The Monin-Obukhov length scale (m) |
| tke_quality | $NT$ | Technical parameter |
| u | $Nz$ x $NT$ | The mean velocity (m s$^{-1}$) component along $x$ |
| v | $Nz$ x $NT$ | The mean velocity (m s$^{-1}$) component along $y$ |
| absU | $Nz$ x $NT$ | The module of the horizontal velocity $U = \left(u^2 + v^2\right)^{1/2}$ |
| gradU | $Nz$ x $NT$ | The vertical gradient of the horizontal velocity $\partial U / \partial z$ |
| absT | $Nz$ x $NT$ | The mean potential temperature (K) profiles given in deviation from an arbitrary constant $\theta_s$. |
| gradT | $Nz$ x $NT$ | The vertical gradient of the potential temperature $\partial \theta / \partial z$ |
| Nbv | $Nz$ x $NT$ | The Brunt-Vaisala frequency (s$^{-1}$), $N = \sqrt{g/\theta_0 \partial \theta / \partial z}$ |
| cs | $Nz$ x $NT$ | The dynamic Smagorinsky constant (non-dimensional), $C_s = l_s / \left(\Delta_x \Delta_y \Delta_z\right)^{1/3}$ where $l_s$ (m) is after Eq. (5) and $\Delta_x, \Delta_y, \Delta_z$ are the run resolution (m) in $(x, y, z)$ directions correspondingly. |
| restau sgstau tau | $Nz$ x $NT$ | The resolved ($\overline{u_i' w'}$), sub-grid $\tau_{ij}$ after Eq. (4) and the total $\tau$ vertical momentum flux (m$^2$ s$^{-2}$). |
| restuw sgstuw uw | $Nz \times NT$ | The same as for $\tau$ but for the stress component ($\overline{u'w'}$). |
| restvw sgstvw vw | $Nz \times NT$ | The same as for $\tau$ but for the stress component ($\overline{v'w'}$). |
| restuv sgstuv uv | $Nz \times NT$ | The same as for $\tau$ but for the stress component ($\overline{u'v'}$). |
| reswt sgswt wt | $Nz \times NT$ | The resolved ($\overline{\theta'w'}$), sub-grid $\tau_{\theta j}$ after Eq. (10) and the total $F_\theta(z)$ vertical temperature flux (K m s$^{-1}$). |
| resut sgsut ut | | The same as for $F_\theta(z)$ but for the flux component ($\overline{u'\theta'}$). |
| resvt sgsvt vt | | The same as for $F_\theta(z)$ but for the flux component ($\overline{v'\theta'}$). |
| tke | $Nz \times NT$ | The total turbulent kinetic energy (m$^2$ s$^{-2}$), $E = 0.5(\sigma_u^2 + \sigma_v^2 + \sigma_w^2)$. |
| resuu sgsuu uu | $Nz \times NT$ | The resolved ($\overline{u'u'}$), sub-grid $L_{11}^L$ after Eq. (6) and the total $\sigma_u^2$ variations of the velocity component $u$ (m$^2$ s$^{-2}$). |
| resvv sgsvv vv | $Nz \times NT$ | The same as for $\sigma_u^2$ but for the velocity component $v$ (m$^2$ s$^{-2}$). |
| resww sgsww ww | $Nz \times NT$ | The same as for $\sigma_u^2$ but for the velocity component $w$ (m$^2$ s$^{-2}$). |
| restt tt | $Nz \times NT$ | The resolved ($\overline{\theta'\theta'}$), and the total $\sigma_\theta^2$ variations of the potential temperature $\theta$ (K$^2$) |
| we | $Nz \times NT$ | The resolved flux of the turbulent kinetic energy, $\overline{w'E}$ (m$^3$ s$^{-3}$) |
| wp | $Nz \times NT$ | The resolved flux of the pressure fluctuations, $\overline{w'p'}$ (m$^3$ s$^{-3}$) |
| wtt | $Nz \times NT$ | The resolved flux of temperature fluctuations, $\overline{w'\theta'\theta'}$ (K$^2$ m s$^{-1}$) |
| pp | $Nz \times NT$ | The resolved scale variations of pressure fluctuations, $\overline{p'p'}$ (m$^4$ s$^{-4}$) |
| iHtau_1 iHtau_5 iHtau_10 | $NT$ | The level number where $\tau$ drops below 1%, 5%, 10% of its surface value $\tau = u_*^2$ |
| rHtau_1 rHtau_5 rHtau_10 | $NT$ | The SBL depth, $h$, defined as the level height where $\tau$ drops below 1%, 5%, 10% of its surface value $\tau = u_*^2$ |
| iHTM | $NT$ | The level number where $Ri_b$ becomes larger than $Ri_c = 0.25$ in the Troen and Mahrt (1986) algorithm. |
| rHTM | $NT$ | The SBL depth, $h$, defined as the level height where $Ri_b$ becomes larger than $Ri_c = 0.25$ in the Troen and Mahrt (1986) algorithm. |
| ldim | 3 | The run dimensions 1 through 3: $Lx, Ly, Lz$ - the domain size (m) in $(x, y, z)$ directions |

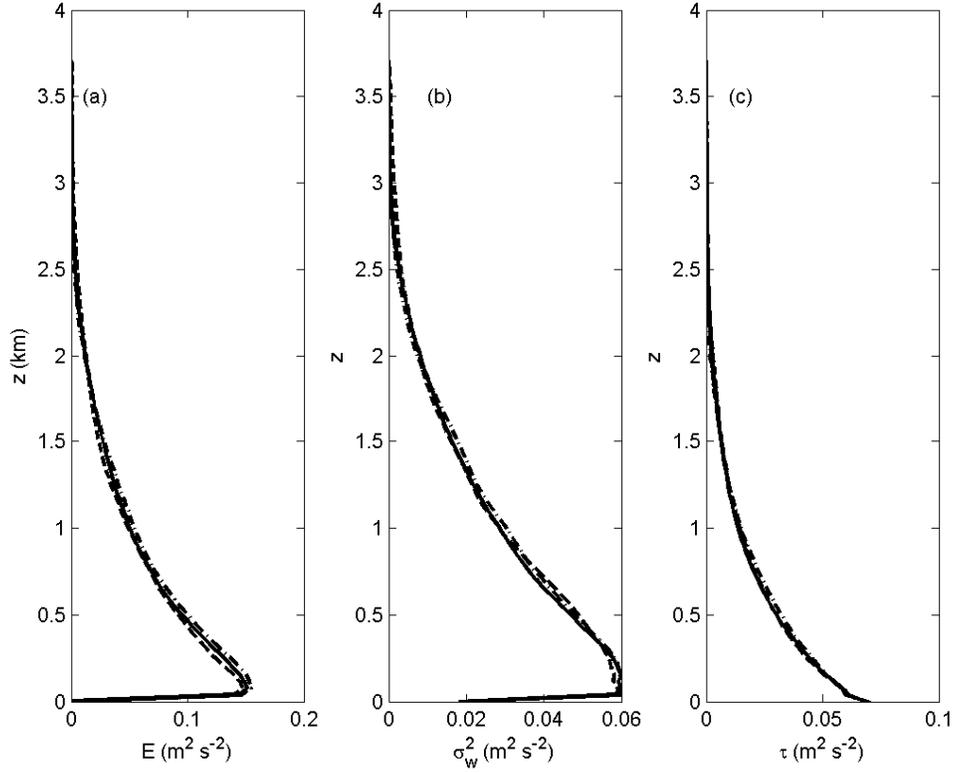

**Figure 1.** Assessment of the LES sensitivity to the horizontal size of the computational domain. The assessment is done for the truly neutral run defined by $U_g = 5$ m s$^{-1}$, $z_0 = 0.1$ m, $\varphi = 45°$N. The domain size varies from 4 x 3 x 3.75 km (dashed curve) through 8 x 6 x 3.75 km (dashed-dot curve) to 16 x 12 x 3.75 km (solid curve). The mesh resolution has been kept the same for all runs as 62.5 m, 46.9 m, 39.1 m correspondingly. Three turbulence statistics are considered: (a) the total turbulent kinetic energy; (b) the resolved vertical velocity component variations; (c) the magnitude of the total vertical momentum flux.

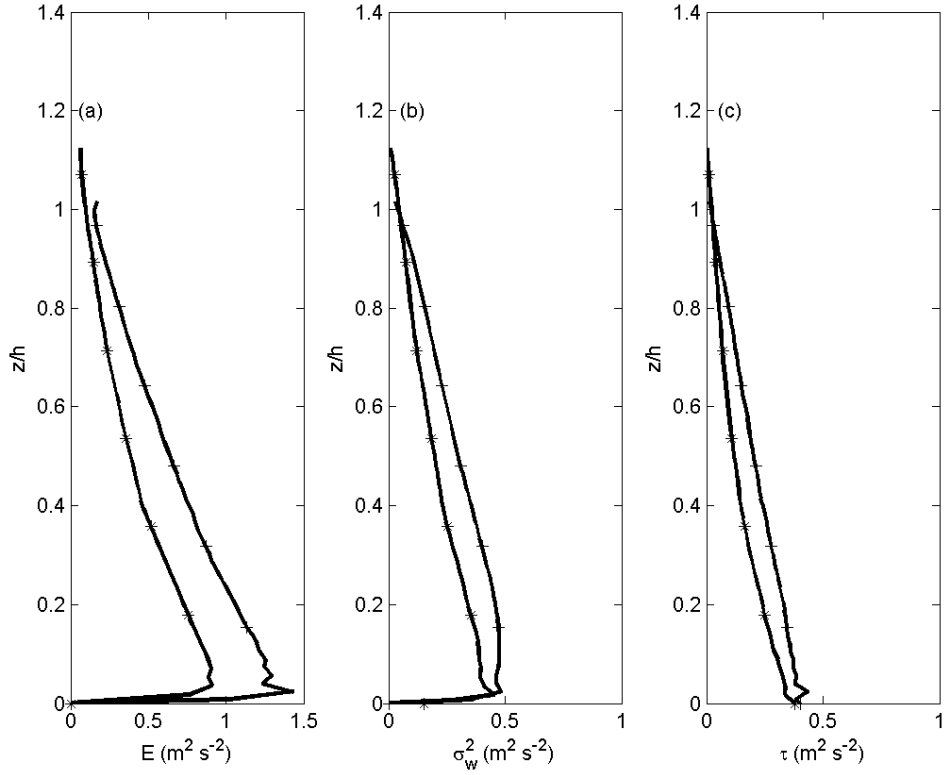

**Figure 2.** Assessment of the LES sensitivity to the vertical size $L_z$ of the computational domain. The assessment is done for the NS SBL runs defined by $U_g = 5$ m s$^{-1}$, $z_0 = 1.0$ m, $F_\theta = -2 \; 10^{-3}$ K m s$^{-1}$, $\varphi = 45°$N. The ratio $h/L_z$ varies from 1 (+) to 0.8 (*). Three turbulence statistics are considered: (a) the total turbulent kinetic energy; (b) the resolved vertical velocity component variations; (c) ) the magnitude of the total vertical momentum flux.

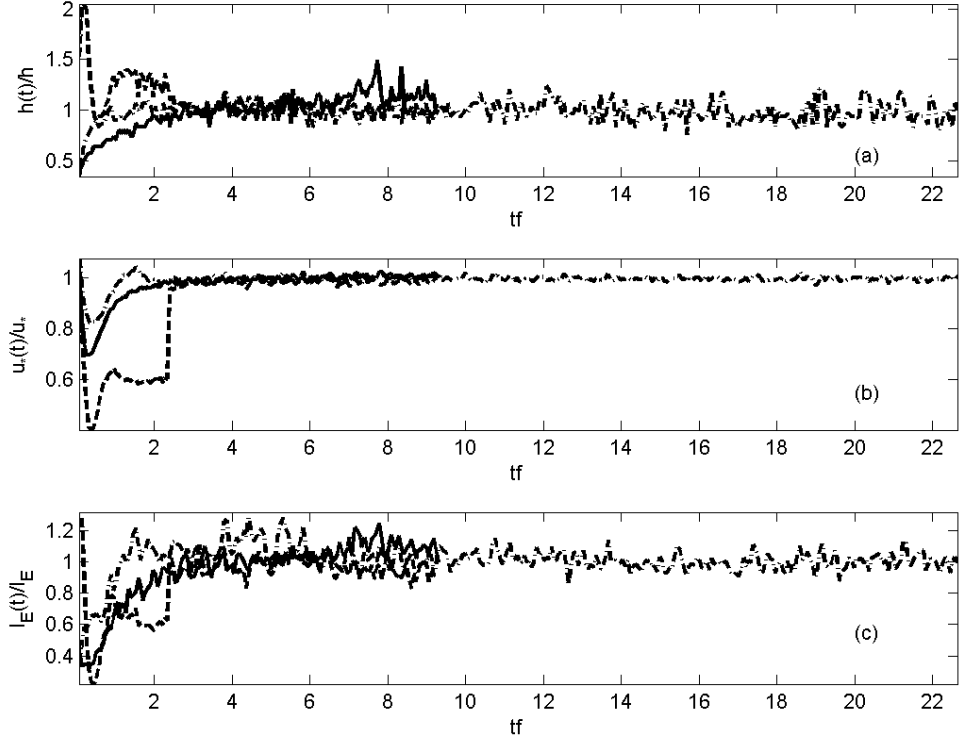

**Figure 3.** Assessment of the equilibrium steady state in the LES runs. The assessment is done for three stably stratified runs defined by $U_g$ = 5 m s$^{-1}$, $F_\theta$ = -10$^{-3}$ K m s$^{-1}$, $z_0$ = 0.1 m, $\varphi$ = 45$^{\rm o}$N and the imposed stability $N$ = 0 s$^{-1}$ (dashed curve), $N$ = 2 10$^{-3}$ s$^{-1}$ (dash-dot curve), and $N$ = 2 10$^{-3}$ s$^{-1}$ (solid curve). Three turbulence statistics are considered: (a) the SBL depth $h(t)$; (b) the surface friction velocity $u_*(t)$; (c) the integral turbulent kinetic energy $I_E(t) = \int_0^{L_z} E(z,t) dz$. All quantities are normalized to their respective values at the maximum $t$ achieved. The time $t$ is normalized by the relevant value of the Coriolis parameter $f$.

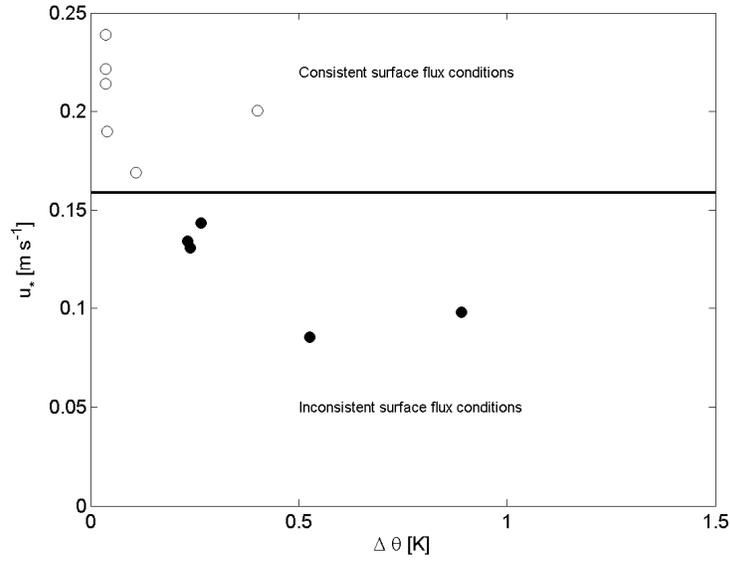

**Figure 5.** Illustration of the consistency analysis of the LES with prescribed surface temperature flux. The analysis was carried out following Basu et al. (2008) for LES runs with $U_g = 5$ m s$^{-1}$, $z_0 = 0.1$ m, $\varphi = 45°$N and different $F_\theta$ and $N$. The black dots signal the physically inconsistent conditions (blacklisted runs).

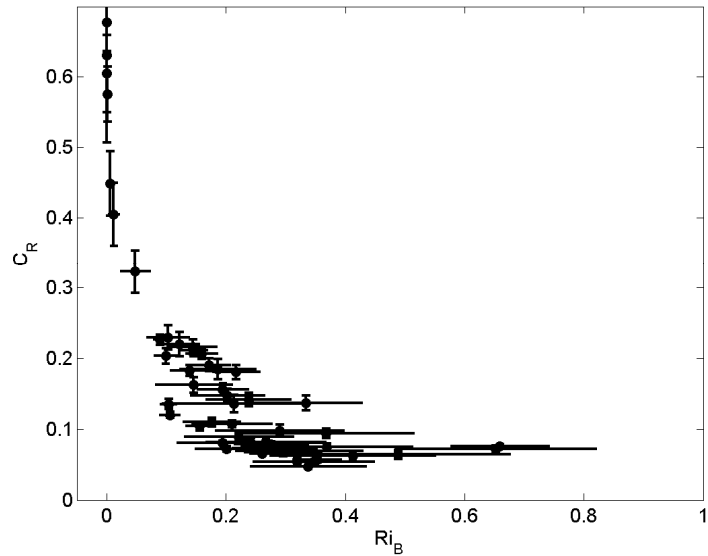

**Figure 6.** The Rossby-Montgomery constant $C_R = h/h_{EBL}$ where the corresponding EBL depth (i.e. in the runs with the same parameters except for $F_\theta = 0$ and $N = 0$; Zilitinkevich et al., 2007) versus the bulk Richardson number. The error bars show one standard deviation.

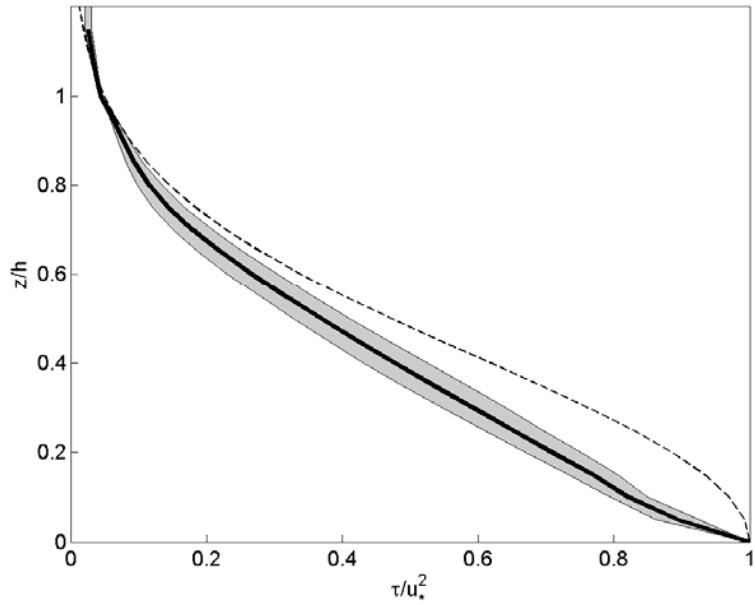

**Figure 7.** The mean inter-run vertical profile (solid line) and one standard deviation intervals (shaded area) of the normalized momentum flux for all runs in DATABASE64. The dashed line is the approximation $\tau/u_*^2 = \exp(-3(z/h)^2)$ from in Zilitinkevich and Esau (2007).

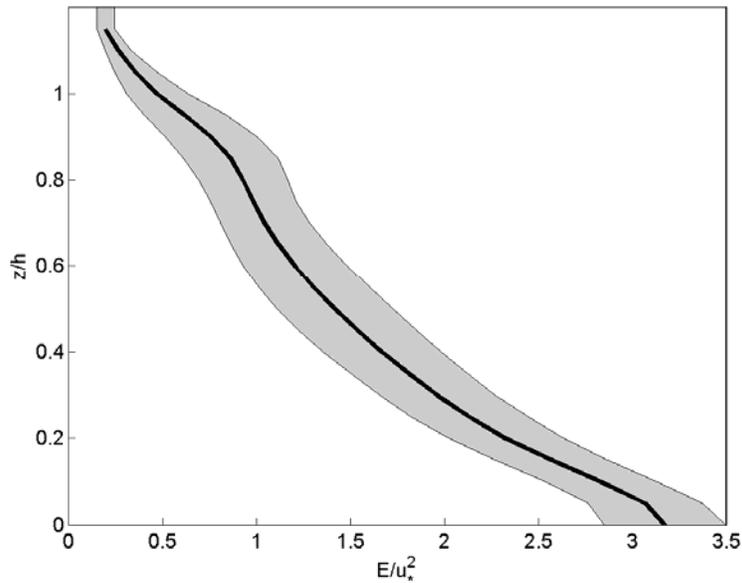

**Figure 8.** The same as in Figure 7 but for the turbulent kinetic energy.

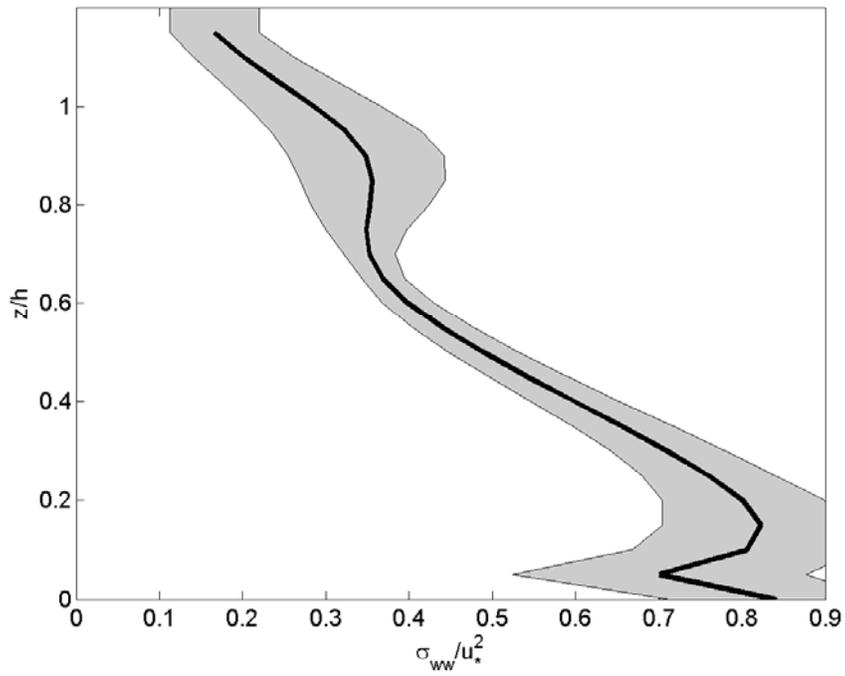

(a)

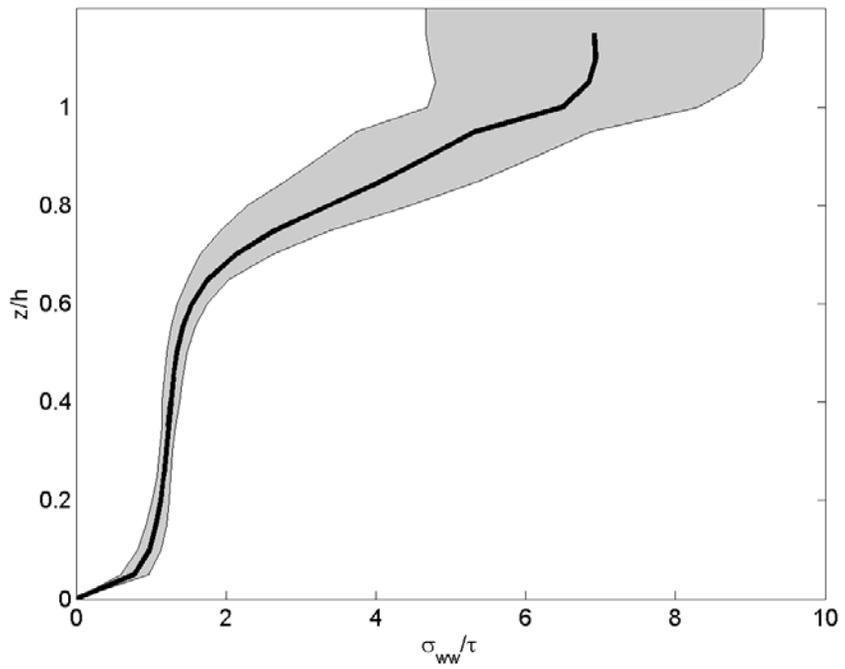

(b)

**Figure 9.** The same as in Figure 7 but for the variations of the vertical component of velocity with (a) the surface scaling, and (b) the local scaling.

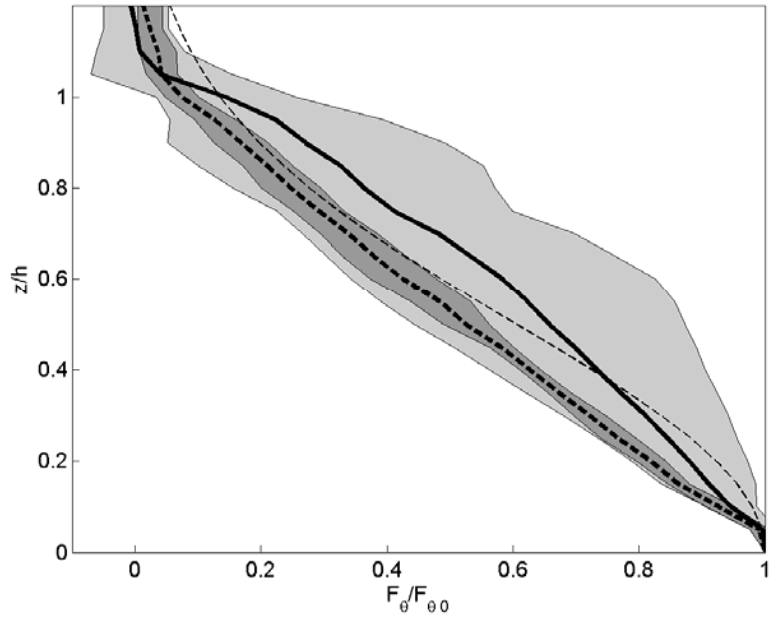

**Figure 10.** The same as in Figure 7 but for the temperature flux. The data for the NS SBL are shown by the bold dashed line and darker shading. The dashed line is the approximation $\tau/u_*^2 = \exp(-2(z/h)^2)$ from Zilitinkevich and Esau (2007).

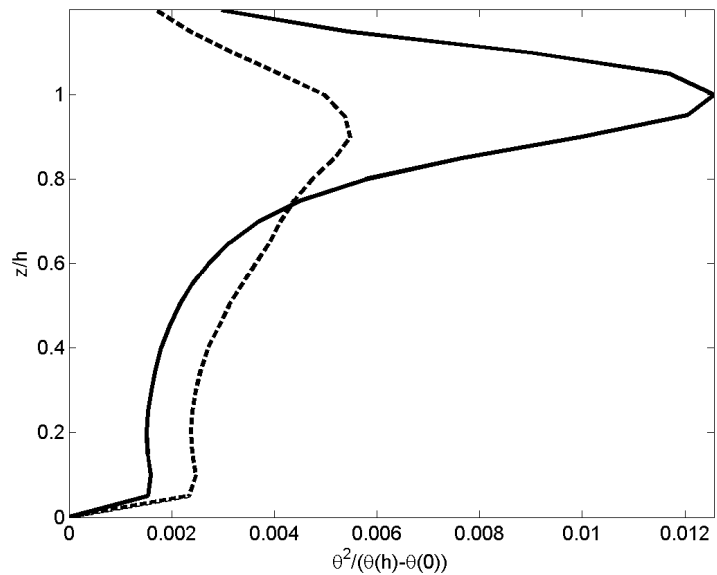

**Figure 11.** The inter-run mean vertical profiles of the temperature fluctuations normalized on the total temperature difference across the SBL in DATABASE64: solid line – the mean over all runs; dashed line – the mean over NS SBL.

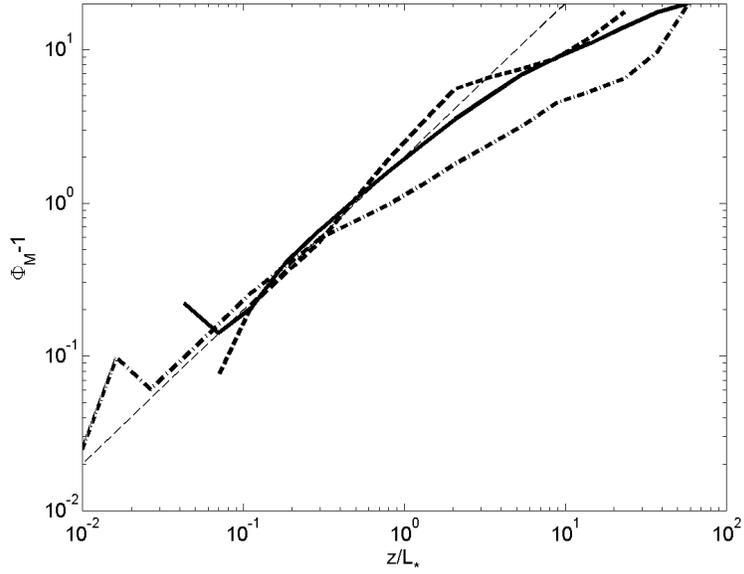

**Figure 12.** The mean inter-run dimensionless velocity gradient $\Phi_M = \kappa z \tau^{-1/2} \partial U / \partial z$ in the SBL ($z < 2/3\ h$) versus the dimensionless height $z/L_*$: solid line – the LS SBL; dashed line – the NS SBL; dash-dotted line – the CN SBL. The thin dashed line is the approximation $\Phi_M = 1 + 2z/L_*$ from Zilitinkevich and Esau (2007).

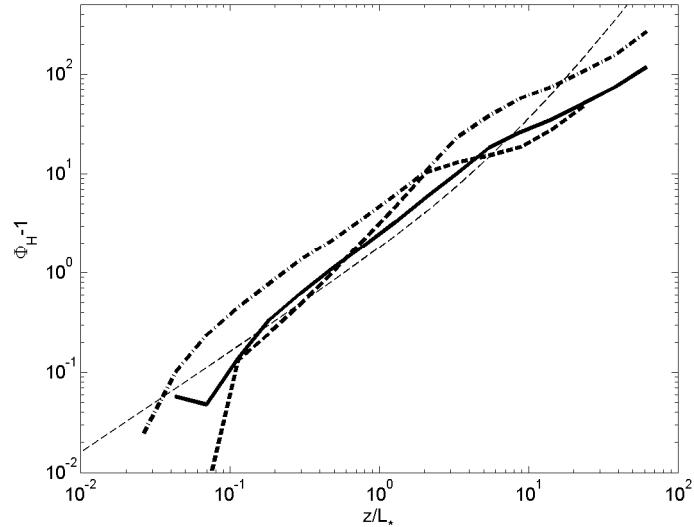

**Figure 13.** The same as in Figure 12 but for $\Phi_H = \kappa_T z \tau^{1/2} F_\theta^{-1} \partial \theta / \partial z$. The thin dashed line is the approximation $\Phi_H = 1 + 1.6 z/L_* + 0.2(z/L_*)^2$ from Zilitinkevich and Esau (2007).

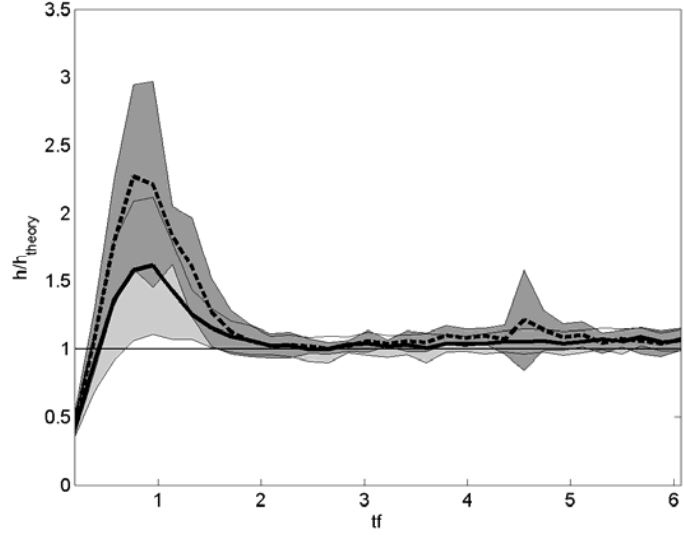

**Figure 14.** The temporal evolution of the normalized SBL depth $h/h_{theory}$ where $h_{theory}$ is theoretical equilibrium depth after Eq. (21). The bold curve and the light shading represent the inter-run mean and the corresponding one standard deviation for all runs in DATABASE64. The dashed curve and the dark shading represent $h/h_{theory}$ only for the NS SBL.

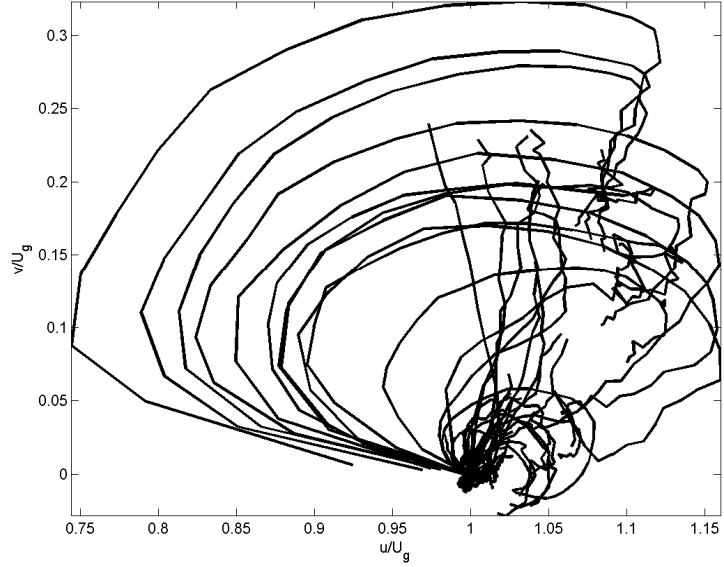

**Figure 15.** The temporal evolution (hodograph) of the normalized velocity $u_i/|U_g|$ at the SBL top, $z = h_{theory}(\infty)$, in DATABASE64. The inertial oscillation does not develop when the initial perturbations do not decay with time, i.e. in the EBL and weakly stable SBL runs.

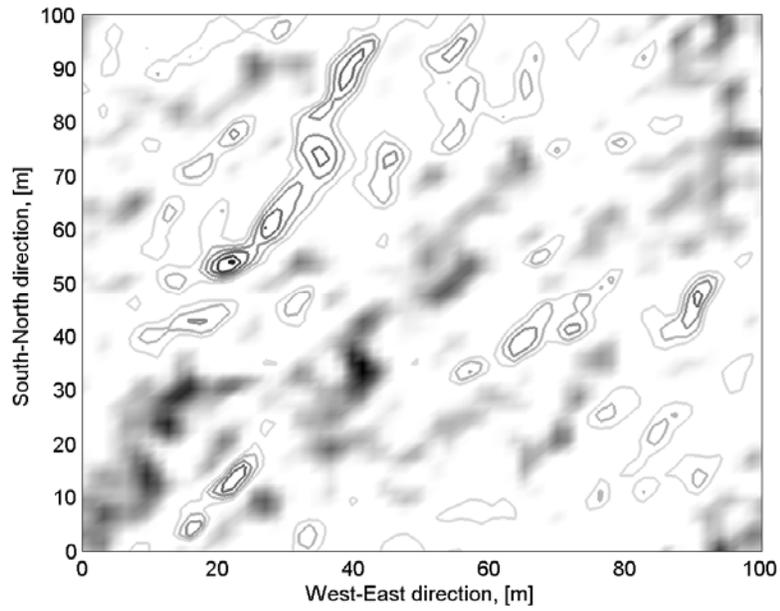

(a)

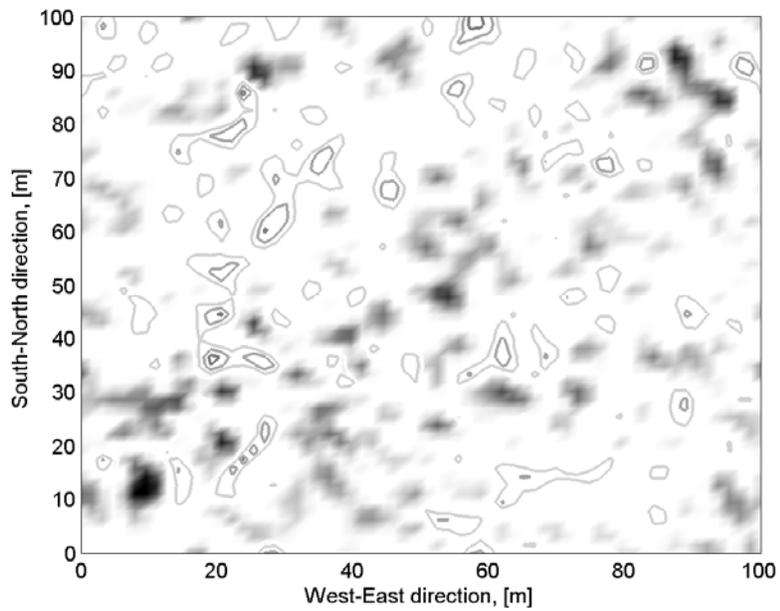

(b)

**Figure 16.** The instant snapshot of the horizontal component of velocity anomalies (a) and the potential temperature anomalies (b) in the most stable SBL in DATABASE64: grey shading – positive anomalies; contours – negative anomalies. The shade and contour interval cover uniformly the range from 10% to 90% of the full range variability.

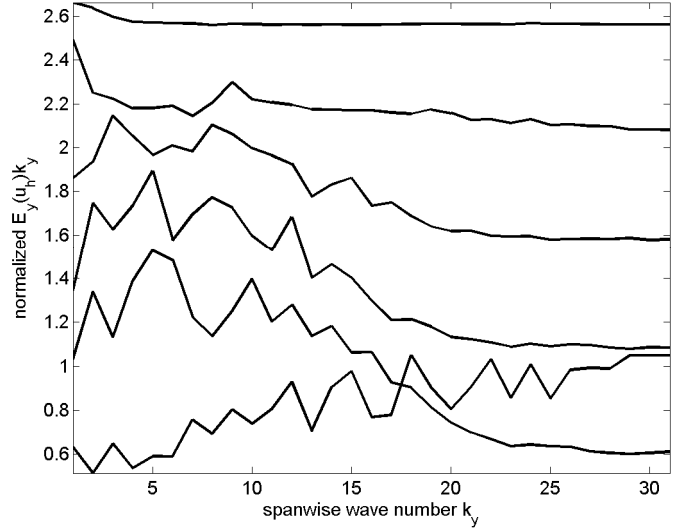

**Figure 17.** The normalized cross-wind spectra of the horizontal velocity at $z = 10\Delta_z, 20\Delta_z, 30\Delta_z, ..., 60\Delta_z$ for the most stable SBL in DATABASE64. The spectra are shifted along ordinate for clarity.

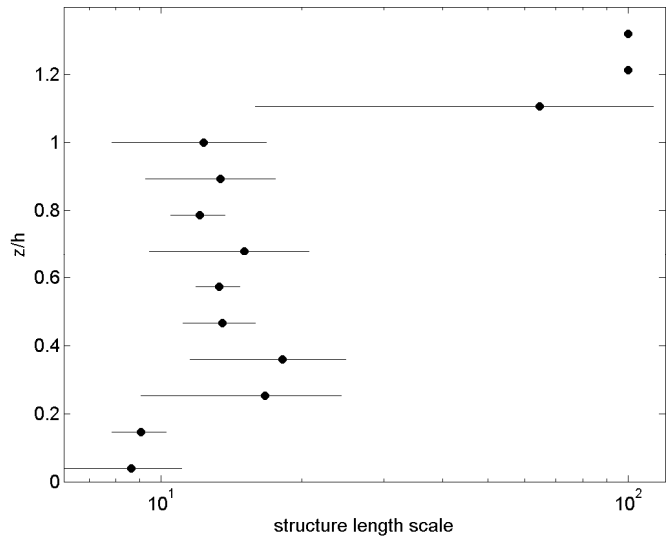

**Figure 18.** The vertical profile of the length scale, $l_y$, of the spectral maximums for the same data as in Figure 17. Three layers are distinguishable: the surface layer $z < 0.3h$ – the length scale increases with the height $l_y \propto z$; the SBL core $0.3h < z < h$ – the length scale is constant $l_y \sim const$; and the free atmosphere $z > h$ where only the largest domain size fluctuations exist.